\documentclass[10pt, twocolumn]{revtex4}
\usepackage[dvipdfmx]{graphicx}
\usepackage{amsmath}
\usepackage{amssymb}

\usepackage{dcolumn}% Align table columns on decimal point
\usepackage{bm}% bold math
\usepackage[normalem]{ulem}
\usepackage{color}
\newcommand{\mcal}[1]{\mathcal{#1}}
\newcommand{\mb}[1]{\mathbb{#1}}

\usepackage{hyperref}

\begin{document}

\preprint{APS/123-QED}

%\title{Phase Transitions in Biological Information Processing under Resource Limitations}
%\title{Phase Transitions induced by Resource Limitations\\ in Biological Information Processing}
\title{Resource Limitations induce Phase Transitions in Biological Information Processing}
% Force line breaks with \\

\author{Takehiro Tottori$^{1,2}$}
%\email{takehiro\_tottori@sat.t.u-tokyo.ac.jp}
\author{Tetsuya J. Kobayashi$^{2}$}
%\altaffiliation[Also at ]{
%Institute of Industrial Science, The University of Tokyo, Tokyo 153-8505, Japan;
%and at Department of Electrical Engineering and Information Systems, Graduate School of Engineering, The~University of Tokyo, Tokyo 113-8654, Japan; 
%and at Universal Biology Institute, The University of Tokyo, Tokyo 113-8654, Japan. 
%}
% \affiliation{
% Department of Mathematical Informatics, Graduate School of Information Science and Technology, The University of Tokyo, Tokyo 113-8654, Japan
% }
\affiliation{
$^{1}$Laboratory for Neural Computation and Adaptation, RIKEN Center for Brain Science, 2-1 Hirosawa, Wako, Saitama 351-0198, Japan\\
$^{2}$Institute of Industrial Science, The University of Tokyo, 4-6-1 Komaba, Meguro, Tokyo 153-8505, Japan
}

\date{\today}

\begin{abstract}
Biological information processing manifests a huge variety in its complexity and capability among different organisms, which presumably stems from the evolutionary optimization under limited computational resources. Starting from the simplest memory-less responsive behaviors, more complicated information processing using internal memory may have developed in the evolution as more resources become available. 
In this letter, we report that optimal information processing strategy can show discontinuous transitions along with the available resources, i.e., reliability of sensing and intrinsic dynamics, or the cost of memory control.
In addition, we show that transition is not always progressive but can be regressed. 
Our result obtained under a minimal setup suggests that the capability and complexity of information processing would be an evolvable trait that can switch back and forth between different strategies and architectures in a punctuated manner. 
% \begin{description}
% Secondary publications and information retrieval purposes.
% You may use the \texttt{description} environment to structure your abstract;
% use the optional argument of the \verb+\item+ command to give the category of each item. 
% \end{description}
\end{abstract}

\maketitle
%\tableofcontents

%%%----------------------------------------------------------------------%%%
{\it Introduction.--}
Biological systems, from bacteria to human beings, possess the ability to process information obtained from the environment to estimate the external world and control themselves for adapting in changing environments.
Recent applications of estimation and control theories revealed that information processing has been optimized not only for human beings \cite{ernst_humans_2002,weiss_motion_2002,kording_bayesian_2004} but even for simple bacteria \cite{bialek_physical_2005,celani_bacterial_2010,mattingly_Escherichia_2021,nakamura_connection_2021,nakamura_optimal_2022}. 
While being optimally tuned, the information processing of different organisms is highly diverse, showing deviations in their complexity and versatility.
As conceived by bounded rationality \cite{simon_rational_1956,lieder_resource-rational_2020,polania_rationality_2024}, such diversity may be rooted in the available resources that the organism can invest for information processing. 
Our brain works by assembling billions of neurons and consuming 20\% of body energy \cite{laughlin_metabolic_1998,lennie_cost_2003,balasubramanian_brain_2021}, whereas \textit{E. coli} optimally senses ligand gradients with a limited number of noisy intracellular reactions by efficiently exploiting tiny amount of energy \cite{elowitz_stochastic_2002,taniguchi_quantifying_2010,govern_optimal_2014}.
The information processing capability, as an evolvable trait, is naturally expected to have evolved in tight relation to the available resources. 
Moreover, the information processing should be affected not only by the resources used directly for operation, i.e., memory size or operational energy, but also by those used implicitly for suppressing stochasticity within cells to realize more reliable sensing and intrinsic dynamics.
Despite its fundamental importance, only a few pioneering works individually addressed this problem \cite{verano_olfactory_2023,sachdeva_optimal_2021,tjalma_trade-offs_2023,govern_energy_2014,lang_thermodynamics_2014,bryant_physical_2023}. We still lack a coherent theory that can integrate various resource limitations and unable to answer a simple question: ``Do optimal information processing strategies progress gradually or punctuatedly along with the expansion of resources?''

This deficiency is partly due to the fact that conventional estimation and control theories \cite{kallianpur_stochastic_1980,chen_bayesian_2003,jazwinski_stochastic_2007,bain_fundamentals_2009,bensoussan_stochastic_1992,yong_stochastic_1999,nisio_stochastic_2015,bensoussan_estimation_2018}, being employed for assessing the optimality of biological information processing, are not designed to accommodate resource limitations. 
In the accompanying paper, we have developed a new theory for biological estimation and control with resource limitations \cite{tottori_theory_2024}. 
By leveraging this theory in this letter, we report that discontinuous phase transitions can occur between memory-dependent information processing and memory-less responsive behaviors along with the availability of different resources, i.e., stochasticity in sensing, stochasticity in intrinsic memory dynamics, and the cost of memory control.
In addition, we show that transition is not always progressive but can be regressed. 
Our result obtained under a minimal setup supports that the capability and complexity of information processing would be an evolvable trait that can switch back and forth between different strategies and architectures in a punctuated manner. 

%%%----------------------------------------------------------------------%%%
{\it Model.--}
\begin{figure*}
	\includegraphics[width=170mm]{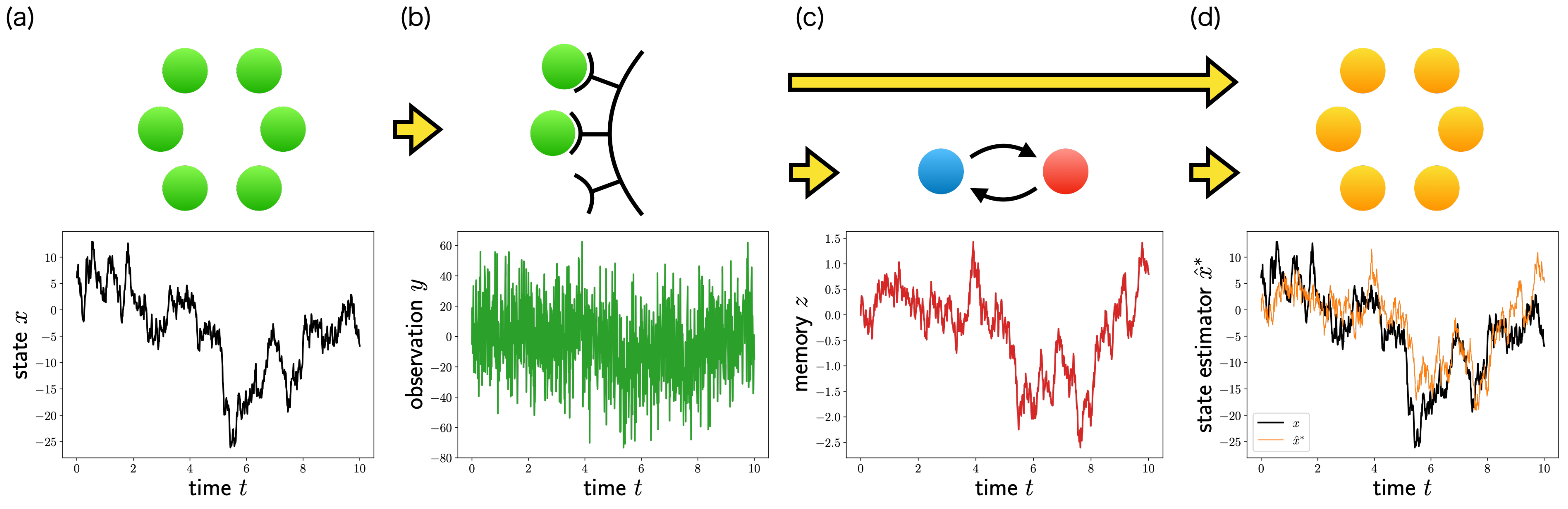}
	\caption{\label{fig: ABIP-diagram}
	Schematic diagram of a minimal model of cellular information processing. 
	(a) Stochastic dynamics of environmental state $x_{t}\in\mb{R}$, e.g., the fluctuation of environmental ligands, that the cell is sensing for adaptation.
    (b) Noisy receptor dynamics $y_{t}\in\mb{R}$, which transmits the environmental information into the cell. 
    (c) Intracellular memory dynamics $z_{t}\in\mb{R}$, which encodes  and accumulates the observed receptor signals for further information processing.
    (d) Downstream molecules exploit both instantaneous receptor observation $y_{t}$ and the memory state $z_{t}$ to generate optimal responses $\hat{x}^{*}_{t}$ (the orange curve) to the current environmental state $x_{t}$ (the black curve). 
	%The sample paths are obtained for $D=100$, $E=500$, $F=1$, $Q=10$, and $M=1$.
    The parameters are $D=100$, $E=500$, $F=1$, $Q=10$, and $M=1$.
	}
\end{figure*}
Estimating environmental states from sensory signals is the simplest but fundamental information processing (Fig. \ref{fig: ABIP-diagram}), with which we start our analysis.
When the sensing is noisy (Fig. \ref{fig: ABIP-diagram}(b)), the past sensing signal should be accumulated for more accurate estimation (Fig. \ref{fig: ABIP-diagram}(c)), which distills into Bayesian sequential computation if optimized without any constraints \cite{kobayashi_implementation_2010,zechner_molecular_2016,mora_physical_2019}. 
However, Bayes's rule requires a deterministic update of the posterior distribution, and the Bayesian posterior distribution becomes an infinite-dimensional function when the environmental state is continuous, meaning that it cannot be faithfully implemented by realistic biological systems such as cells with intrinsic stochasticity and finite memory size. 
Moreover, complex computation may require additional energy expenditure. 
Thus, when the available resources are limited, signal processing with memory may not necessarily be advantageous than memory-less responsive behaviors that use only instantaneous observations to react to the environment. 

To address this situation under the minimal setup, we assume that $x_{t}\in\mb{R}$ represents an environmental state such as ligand concentration, which evolves by the following Ornstein-Uhlenbeck process (Fig. \ref{fig: ABIP-diagram}(a)): 
\begin{align}
	dx_{t}&=-x_{t}dt+\sqrt{D}d\omega_{t},
	\label{eq: ABIP-state SDE}
\end{align}
where $\omega_{t}\in\mb{R}$ is a standard Wiener process. $D\geq0$ is the intensity of the state noise. 
The average of $x_{t}$ is rescaled to $0$ without loosing generality.
The noisy observation of the state by a sensory system such as chemical receptors is represented by $y_{t}\in\mb{R}$ (Fig. \ref{fig: ABIP-diagram}(b)), which follows Gaussian distribution with mean $x_{t}$ and variance $E\geq0$: 
\begin{align}
	y_{t}\sim\mcal{N}(y_{t}|x_{t},E).
	\label{eq: ABIP-observation SDE}
\end{align}
In the usual Bayesian estimation framework, we derive the dynamics of posterior probability density function $p(x_{t}|y_{t},...,y_{0})$ of environmental state $x_{t}$ given observation history $y_{t},...,y_{0}$.  
To account for resource limitations, e.g., intrinsic noise, or finite memory size, in our framework, we explicitly consider the intracellular memory state $z_{t}\in\mb{R}$ whose dynamics are modeled by the following stochastic differential equation (SDE) (Fig. \ref{fig: ABIP-diagram}(c)): 
\begin{align}
	dz_{t}&=v(y_{t},z_{t}) dt+\sqrt{F}d\xi_{t}, 
	\label{eq: ABIP-memory SDE}
\end{align}
where standard Wiener process $\xi_{t}\in\mb{R}$ represents the intrinsic noise with intensity $F\geq0$. 
The dynamics of $z_{t}$ is regulated by observation $y_{t}$ and self-feedback as $v(y_{t},z_{t})\in\mb{R}$.
Then, we derive the optimal regulation $v(y_{t},z_{t})$ under resource constraints by minimizing the objective function
\begin{align}
	&J[v, \hat{x}]:=\lim_{T\to\infty}\frac{1}{T}\mb{E}\left[\int_{0}^{T}\left(Mv_{t}^{2} + Q(x_{t}-\hat{x}_{t})^{2}\right)dt\right],
	\label{eq: ABIP-OF}
\end{align}
where $M>0$ and $Q>0$. The first term is the quadratic regulation cost of the memory dynamics, which penalizes costly dynamics. The second term is the estimation error of the environmental state $x_{t}$ by using the observation and memory state at $t$ where $\hat{x}_{t}=\hat{x}(y_{t},z_{t})$ is an auxiliary function converting the observation and memory state into the estimate of $x_{t}$. This function accounts for the fact that the information stored in intracellular memory has to be decoded appropriately before being exploited at the downstream (Fig. \ref{fig: ABIP-diagram}(d)).
The optimal memory control function $v^{*}$ is obtained together with the optimal state estimator function $\hat{x}^{*}$ by solving the joint optimization problem:
\begin{align}
	\hat{x}^{*},v^{*}:=\arg\min_{\hat{x},v}J[\hat{x},v]. \label{eq:joint_opt}
\end{align}

%%%----------------------------------------------------------------------%%%
{\it Optimal solution.--}
While the formulation is straightforward, solving this optimization problem is technically non-trivial, which is elaborated in the accompanying paper. 
Under our setup, optimal $v^{*}$ and $\hat{x}^{*}$ are obtained semi-analytically as
\begin{align}
	v^{*}(y,z)&=-M^{-1}\Pi_{zx}\hat{x}^{*}(y,z)-M^{-1}\Pi_{zz}z,\label{eq: ABIP-optimal memory control}\\
	\hat{x}^{*}(y,z)&=\mb{E}_{p(x|y,z)}[x]=K_{xy}y+K_{xz}z,\label{eq: ABIP-optimal state estimator}
\end{align}
where the values of control gains, $\Pi_{zx}$ and $\Pi_{zz}$, and estimation gains, $K_{xy}$ and $K_{xz}$, are nonlinear functions of $D$, $E$, $F$, $M$, and $Q$. 
For a given set of these parameters, the gains are obtained by jointly solving the Hamilton-Jacobi-Bellman (HJB) and Fokker-Planck (FP) equations (see the accompanying paper) \cite{tottori_theory_2024}.
Figure \ref{fig: ABIP-diagram}(c) shows the optimal memory dynamics under optimal regulation.
In addition, optimal $\hat{x}^{*}(y,z)$ becomes identical to $\mb{E}_{p(x|y,z)}[x]$, which is consistent with statistical estimation theory and demonstrated in Fig. \ref{fig: ABIP-diagram}(d). 

It should be noted that our theory can cover general nonlinear dynamics for  $x_{t}$ and non-quadratic function for the objective $J[v, \hat{x}]$, which results in nonlinear $v^{*}(y_{t},z_{t})$ and $\hat{x}^{*}(y_{t},z_{t})$. 
Nevertheless, we work on the linear dynamics for $x_{t}$ [Eq. (\ref{eq: ABIP-state SDE})] and quadratic objective function [Eq. (\ref{eq: ABIP-OF})] in this letter to prove that the transitions we are going to observe purely stems from the nature of resource-limited optimal estimation by excluding any unnecessary nonlinearities that can be a potential source of transition-like behaviors. 
Moreover, this setup allows us to investigate various parameters in the form of a phase diagram owing to efficient computations of the optimization problem, which is computationally prohibitive under a general setup.  

\begin{figure}
	\includegraphics[width=85mm]{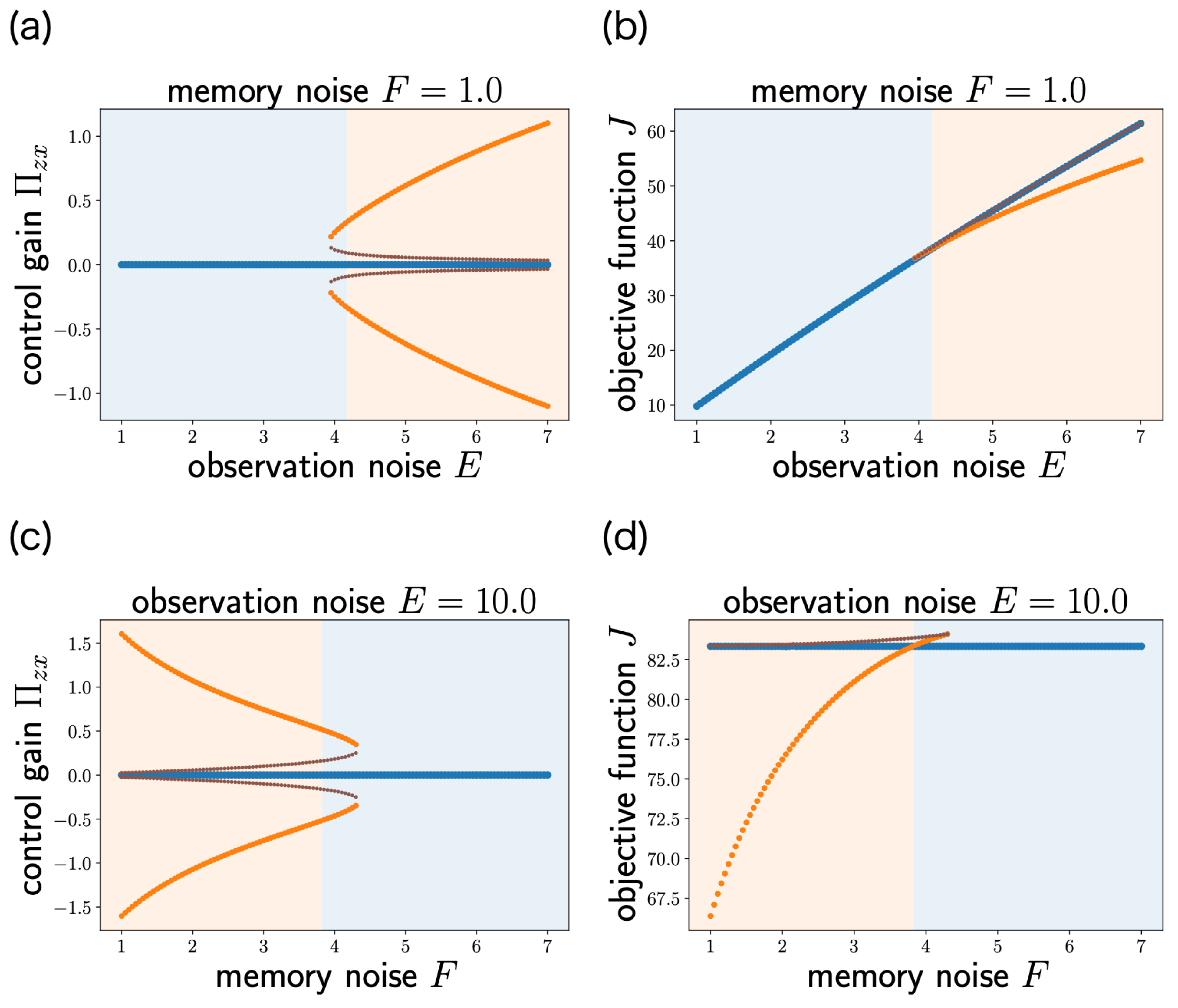}
	\caption{\label{fig:transition}
	Phase transitions between optimal estimation strategies for control gain $\Pi_{zx}$ [Eq. (\ref{eq: ABIP-optimal memory control})] and objective function $J[v, \hat{x}]$ as functions of $E$ (a,b) and $F$ (c,d).
    Blue, brown, and orange dots correspond to the values for the strategies satisfying the stationarity condition of $J[v, \hat{x}]$.
    The responsive strategy is optimal in the blue region whereas the memory-dependent strategy is optimal in the orange region. 
    The memory-dependent strategy (orange dots) appears discontinuously in the blue region where it is suboptimal (the responsive one is still optimal) and the optimality switches at the boundary of the blue and orange regions. 
	The rest of the parameters are $D=100$, $Q=10$, and $M=1$.
	}
\end{figure}
\begin{figure}
	\includegraphics[width=85mm]{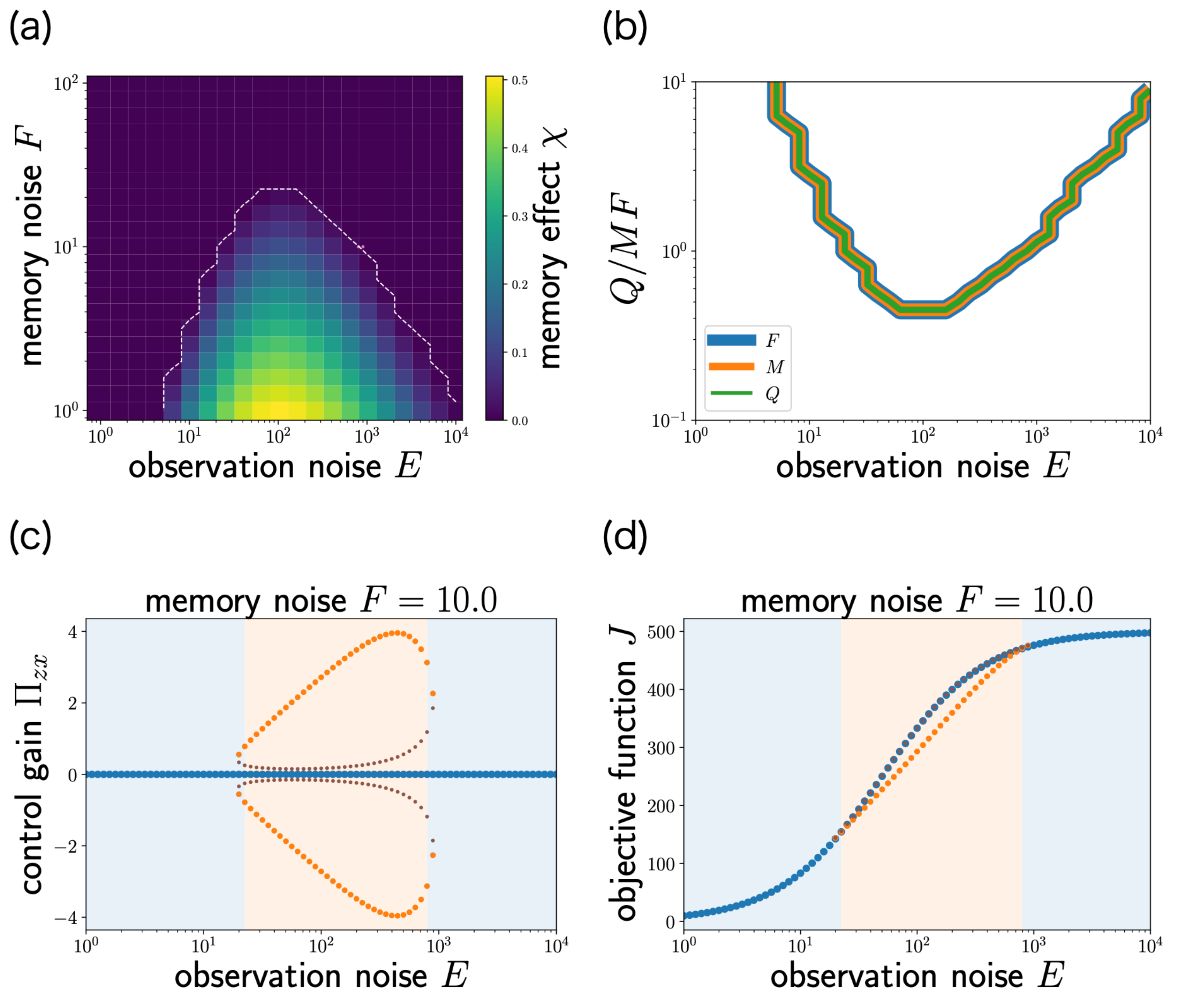}
	\caption{\label{fig:diagram}
    (a) A phase diagram of memory effect $\chi$ for observation noise $E$ and intrinsic noise $F$. Memory effect $\chi$, defined by $(J_{\rm without\ memory}^{*}-J_{\rm with\ memory}^{*})/J_{\rm without\ memory}^{*}$, quantifies the relative contribution of memory. 
    $J_{\rm with/without\ memory}^{*}$ is the minimum value of the objective function with/without memory. 
    (b) Phase boundaries as a function of $E$ and $Q/MF$. 
    Blue, orange, and green curves correspond to the cases where $F$, $M$, and $Q$ are varied, respectively.  
    (c,d) Control gain $\Pi_{zx}$ (c) and objective function $J$ (d) as a function of $E$ for $F=10$, demonstrating the forward and reverse transitions between responsive and memory-dependent strategies. 
    The color codes and the rest of the parameters are the same as in Fig. \ref{fig:transition}. 
	}
\end{figure}

%%%----------------------------------------------------------------------%%%
{\it Phase Transition.--}
We investigate the dependency of $v^{*}(y,z)$ on resource availability, e.g., observation noise $E$, intrinsic noise $F$, and regulatory cost $M$ by focusing on optimal control gain $\Pi_{zx}$ in Eq. (\ref{eq: ABIP-optimal memory control}).
If $\Pi_{zx}=0$, $v^{*}(y,z)$ does not depend on $y$, meaning that memory $z_{t}$ does not accumulate past observations at all. 
Thus, the associated $\hat{x}^{*}(y,z)$ reduces to be responsive, i.e., the optimal estimation of the environment $\hat{x}^{*}(y,z)$ depends only on instantaneous observation $y_{t}$.
If $\Pi_{zx}\neq 0$, $z_{t}$ accumulates past observations and the optimal $\hat{x}^{*}(y,z)$ integrates observation $y_{t}$ and memory $z_{t}$ to estimate the environment.

Figures \ref{fig:transition}(a) and (b) show the control gain and the optimal value of the objective function as functions of observation noise $E$. 
When  $E$ is small, i.e., the observation is reliable, $\Pi_{zx}=0$ and the responsive dynamics becomes optimal. 
As the observation becomes less reliable ($E$ increases), the locally optimal solution with finite $\Pi_{zx}\neq 0$ emerges discontinuously while the responsive behavior, i.e., $\Pi_{zx}=0$, remains the globally optimal strategy. 
As $E$ increases further, global optimality switches from the responsive strategy to memory-dependent one while the responsive still survives as a locally optimal but globally suboptimal solution.
We also have observed similar behaviors for the cases where memory $z_{t}$ becomes more reliable or less costly, i.e., when the intrinsic stochasticity $F$ of $z_{t}$ is decreased (Fig. \ref{fig:transition} (c,d)) or when the memory control cost $M$ is decreased (see the accompanying paper) \cite{tottori_theory_2024}. 
This result, obtained under the setup with minimal nonlinearity, indicates that discontinuity fundamentally underlies the transition of the optimal information processing strategies as a function of resources.
The discontinuity qualitatively impacts the evolutionary scenario of information processing because small perturbations of the control gain around $\Pi_{zx}=0$ via mutation cannot generate advantageous mutants even if the responsive dynamics is no longer optimal.
To overcome the gap between suboptimal and optimal strategies, a large mutational perturbation or systematic accumulation of mutations is required, which would make the evolution of information processing strategies punctuated rather than gradual as the change in the available resources.

To scrutinize the interrelation between different resource-related parameters, we computed the phase diagram for observation noise $E$ and intrinsic noise $F$ (Fig. \ref{fig:diagram}(a)).
From the diagram, we identify the critical intrinsic noise intensity $F^{\dagger}$ above which memory usage is no longer encouraged for any $E$. 
In addition, for a fixed intensity of intrinsic noise $F<F^{\dagger}$, the transition from responsive to memory-dependent strategies is accompanied by its reverse transition as the observation noise intensity $E$ increases (Fig. \ref{fig:diagram}(c,d)), meaning that memory is advantageous only when the observation is moderately noisy.
Furthermore, the optimal observation noise $E^{*}(F)$ exists for fixed $F<F^{\dagger}$ where the performance gain by exploiting memory peaked (Fig. \ref{fig:diagram}(a)). 
This non-monotonous dependency appears not only for observation noise $E$ but also for state noise $D$, while the memory effect $\chi$ changes monotonously with respect to intrinsic noise $F$, regulatory cost $M$, and estimation error $Q$ (see the accompanying paper) \cite{tottori_theory_2024}. 
Thus, extrinsic noise such as observation and state noise may contribute to evolution of information processing in a more complicated manner than intrinsic noise and cost parameters by inducing both progressive and regressive transitions.
We also find that the phase boundaries for $F$, $M$, and $Q$ are nearly identical when plotted against $Q/MF$ (Fig. \ref{fig:diagram}(b)), suggesting that increasing the weight of estimation error $Q$ is nearly equivalent to decreasing control cost $M$ and intrinsic noise $F$ in terms of phase transitions. 

These phenomena cannot be observed within the Bayesian theory, where the use of Bayesian updates always improves estimation performance. 
This aligns with our results because the setup of Bayesian theory corresponds to the limit of no intrinsic noise, $F \to 0$, with no memory control cost $M \to 0$ (equivalently $Q \to \infty$).
Thus, the conventional Bayesian theory addresses only limited and ideal situations of biological information processing, overlooking the qualitative impacts of resource limitation. 

\begin{figure}
	\includegraphics[width=85mm]{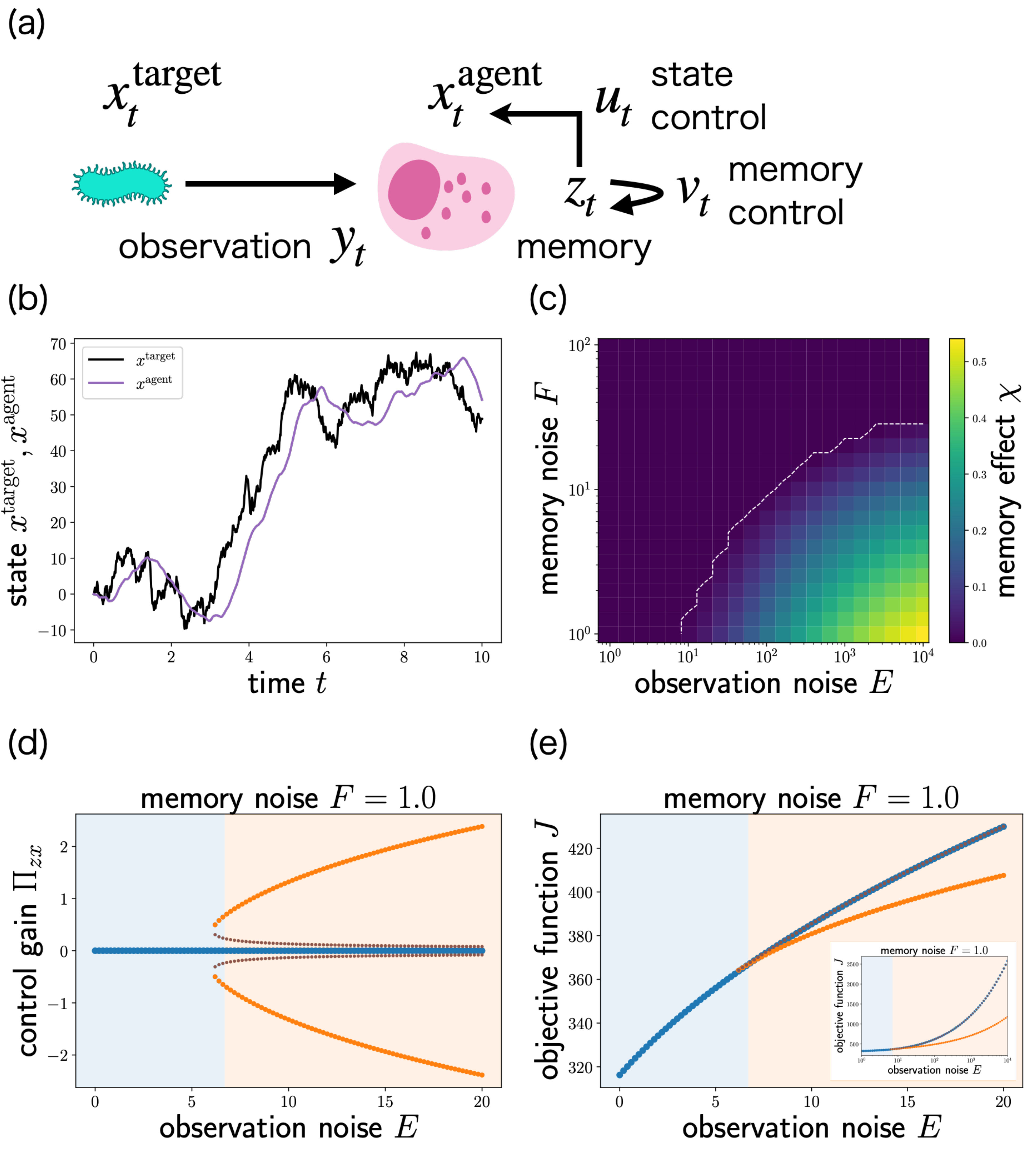}
	\caption{\label{fig: ABDM-diagram}
	(a) Schematic diagram of a target tracking problem. An agent, e.g., a cell, estimates the position of the target $x_{t}^{\rm target}$ from noisy observation $y_{t}$ and  memory $z_{t}$ and tracks it by controlling its own position $x_{t}^{\rm agent}$ through state control $u_{t}$. 
    The memory control $v_{t}$ is jointly optimized with $u_{t}$ to compress past observation history $y_{t-dt},...,y_{0}$ into memory $z_{t}$ so that it provides the relevant information for state control. 
	(b) The trajectories of the target $x_{t}^{\rm target}$ and the agent $x_{t}^{\rm agent}$ for $E=500$ and $F=1$.
    (c) Phase diagram of memory effect $\chi$ as function of observation noise $E$ and intrinsic noise $F$. 
    Memory effect $\chi$ is defined as in Fig. \ref{fig:diagram}.
    (d,e) Phase transitions for control gain $\Pi_{zx}$ (d) and the objective function (e) as functions of $E$. The color codes are the same as in Fig. \ref{fig:transition}.
	The rest of the parameters are $D=100$, $Q=10$, $R=1$, and $M=1$.
	}
\end{figure}

%%%----------------------------------------------------------------------%%%
{\it Extension to include control.--}
We further extend our analysis to a biological control problem in which a cell controls itself to chase a target bacteria based on stochastic observation (Fig. \ref{fig: ABDM-diagram}(a)). 
The target tracking problem is prevalent in many biological phenomena, such as cellular chemotaxis \cite{krummel_t_2016,nakamura_optimal_2022,sowinski_semantic_2023,rode_information_2024}, insect olfactory search \cite{heinonen_optimal_2023,verano_olfactory_2023,celani_olfactory_2024}, and animal foraging behaviors \cite{viswanathan_optimizing_1999,benichou_optimal_2005,viswanathan_levy_2008,benichou_intermittent_2011,hein_natural_2016}. 
While the conventional Bayesian estimation theory and optimal control theory are not seamlessly integrated, our theory with resource limitation can straightforwardly formulate both estimation and control problems (see the accompanying paper for more details) \cite{tottori_memory-limited_2022,tottori_forward-backward_2023,tottori_decentralized_2023,tottori_theory_2024}. 
We show that the discontinuous phase transitions also appear in this target tracking problem. 

%%%----------------------------------------------------------------------%%%

Let $x_{t}^{\rm target}\in\mb{R}$ be the target position and $x_{t}^{\rm agent}\in\mb{R}$ be that of the agent (a cell), which evolves by the following diffusion process: 
\begin{align}
	dx_{t}^{\rm target}=\sqrt{D}d\omega_{t},\quad dx_{t}^{\rm agent}&=-u_{t}dt,\label{eq: ABDM-agent state SDE}
\end{align}
where $\omega_{t}\in\mb{R}$ is a standard Wiener process, $D\geq0$ is the diffusion constant, and $u_{t}\in\mb{R}$ is the state control of the agent.
From the translation invariance, the target tracking problem concerns only with the difference $x_{t}:=x_{t}^{\rm target}-x_{t}^{\rm agent}$, which evolves by the following SDE: 
\begin{align}
	dx_{t}&=u_{t}dt+\sqrt{D}d\omega_{t}.\label{eq: ABDM-state SDE}
\end{align}
Let the observation of the agent be $y_{t}\in\mb{R}$, which is generated from the Gaussian distribution with mean $x_{t}$: $y_{t}\sim\mcal{N}(y_{t}|x_{t},E)$ where $E\geq0$ is the intensity of the observation noise. 
We model the intrinsic dynamics of the agent memory $z_{t}\in\mb{R}$ by the following SDE: $	dz_{t}=v_{t}dt+\sqrt{F}d\xi_{t}$, where $v_{t}\in\mb{R}$ is the memory control of the agent, $\xi_{t}\in\mb{R}$ is a standard Wiener process, and $F\geq0$ is the intrinsic noise intensity of the memory dynamics. 
The objective function to be minimized is
\begin{align}
	&J[u,v]:=\lim_{T\to\infty}\frac{1}{T}\mb{E}\left[\int_{0}^{T}\left(Qx_{t}^{2}+Ru_{t}^{2}+Mv_{t}^{2}\right)dt\right], \nonumber
\end{align}
where the three terms are the distance between the target and the agent, the state control cost, and the memory control cost with weighting parameters, $Q>0$, $R>0$, and $M>0$, respectively. 

Optimal state control $u^{*}_{t}$ and memory control $v^{*}_{t}$ are obtained via minimization $u^{*},v^{*}:=\arg\min_{u,v}J[u,v]$, which results in 
\begin{align}
	u^{*}(y,z)&=-R^{-1}\Pi_{xx}\hat{x}^{*}(y,z)-R^{-1}\Pi_{xz}z,\label{eq: ABDM-optimal state control}\\
	v^{*}(y,z)&=-M^{-1}\Pi_{zx}\hat{x}^{*}(y,z)-M^{-1}\Pi_{zz}z.\label{eq: ABDM-optimal memory control}
\end{align}
where $\hat{x}^{*}(y,z)$ is as in Eq. (\ref{eq: ABIP-optimal state estimator}). 
Control gains, $\Pi_{xx}$, $\Pi_{xz}$, $\Pi_{zx}$, and $\Pi_{zz}$ are obtained from HJB and FP equations (see accompanying paper for more details) \cite{tottori_theory_2024}.

Figure \ref{fig: ABDM-diagram}(b) demonstrates that the agent tracks the target accurately despite the high observation noise. 
Similarly to the estimation problem, we observe the discontinuous phase transitions from $\Pi_{zx}=0$ to $\Pi_{zx}\neq 0$ along with the increase in the observation noise $E$ (Fig. \ref{fig: ABDM-diagram}(d,e)) or the decrease in the memory noise $F$ (see the accompanying paper) \cite{tottori_theory_2024}. 
This means that there exists a transition from the responsive control where only instantaneous observation $y_{t}$ is used for state control to the memory-dependent control where memory $z_{t}$ is additionally exploited for the state control.

Finally, we computed the phase diagram for observation noise $E$ and intrinsic noise $F$, in which no reverse transition was observed even if $E$ is increased (Fig. \ref{fig: ABDM-diagram}(c)). 
This means that using memory for state control is still advantageous even if the observation becomes extremely noisy. 
This differentiation would be attributed to the unbounded state variances in the target tracking problem; even a piece of subtle information for $x_{t}$ could benefit for suppressing the divergence of $x_{t}$ to $\pm \infty$, which accompanies the divergence of $J$. 
We also find that the scaling relationship, $Q/MF$, does not hold in the target tracking problem (see the accompanying paper) \cite{tottori_theory_2024}. 
These results indicate that the non-monotonicity and the scaling relationship are model-dependent, while the discontinuity is a fundamental characteristic under resource limitations. 
Overall, our analysis spotlights the potential diversity and complexity in the optimal ways to employ memory for estimation and control under resource limitations. 

{\it Discussion.--}
In this work, we have identified discontinuous phase transitions in memory dynamics for optimal estimation and control as functions of available resources. 
Because the accumulation and processing of information using memory are fundamental for all intelligent operations, our theory could be a theoretical basis for understanding the development and evolution of intelligent biological agents. 

While we investigated the simplest linear environmental dynamics with the quadratic objective function, our general theory can accommodate nonlinear dynamics and non-quadratic objectives. 
In principle, it allows us to investigate optimal information processing for more complex environmental dynamics such as oscillations under more thermodynamically realistic costs \cite{govern_energy_2014,lang_thermodynamics_2014,bryant_physical_2023}.
Moreover, by considering imbalanced dimensions of the environmental and memory states, we can address the problem of information compression via the internal state, which is a dynamic version of static information compression in machine learning \cite{bishop_pattern_2006,van_der_maaten_dimensionality_2007,lee_nonlinear_2007}. 
For all these applications, more efficient and versatile methods are required to solve the joint optimization problem [Eq. (\ref{eq:joint_opt})]. 
We may employ various methods developed in computational physics for addressing this technical issue.

The first author received a JSPS Research Fellowship (Grant Number 22KJ0557). 
This research was supported by JST CREST (Grant Number JPMJCR2011) and JSPS KAKENHI (Grant Number 19H05799, 24H02148, and 24H01465).

%%%----------------------------------------------------------------------%%%
\bibliography{230601_MPT_ref}
\end{document}